\documentclass[prl, 11pt,letterpaper,superscriptaddress,floatfix,footinbib,notitlepage]{revtex4-1}

\usepackage{textcomp}
\usepackage{amsmath}
\usepackage{amsfonts}
\usepackage{amssymb}
\usepackage{graphicx}
\usepackage{siunitx}
\usepackage{caption}
\usepackage{setspace}
\setcitestyle{super}
\usepackage[colorlinks, citecolor={blue}, linkcolor={black}]{hyperref}
\begin{document}
\raggedbottom

\author{Daniel Rodan-Legrain}
\thanks{These authors contributed equally}
\affiliation{Department of Physics, Massachusetts Institute of Technology, Cambridge, Massachusetts 02139, USA}

\author{Yuan Cao}
\thanks{These authors contributed equally}
\email{caoyuan@mit.edu}
\affiliation{Department of Physics, Massachusetts Institute of Technology, Cambridge, Massachusetts 02139, USA}

\author{Jeong Min Park}
\thanks{These authors contributed equally}
\affiliation{Department of Physics, Massachusetts Institute of Technology, Cambridge, Massachusetts 02139, USA}

\author{Sergio de la Barrera}
\affiliation{Department of Physics, Massachusetts Institute of Technology, Cambridge, Massachusetts 02139, USA}

\author{Mallika T. Randeria}
\affiliation{Department of Physics, Massachusetts Institute of Technology, Cambridge, Massachusetts 02139, USA}

\author{Kenji Watanabe}
\author{Takashi Taniguchi}
\affiliation{National Institute for Materials Science, Namiki 1-1, Tsukuba, Ibaraki 305-0044, Japan}
\author{Pablo Jarillo-Herrero}
\email{pjarillo@mit.edu}
\affiliation{Department of Physics, Massachusetts Institute of Technology, Cambridge, Massachusetts 02139, USA}

\title{
Highly Tunable Junctions and Nonlocal Josephson Effect in Magic Angle Graphene Tunneling Devices
}

\date{\today}

\maketitle

\textbf{
Magic-angle twisted bilayer graphene (MATBG) has recently emerged as a highly tunable two-dimensional (2D) material platform exhibiting a wide range of phases, such as metal, insulator, and superconductor states.  \cite{cao_correlated_2018, cao_unconventional_2018, yankowitz_tuning_2019, lu_superconductors_2019}. Local electrostatic control over these phases may enable the creation of versatile quantum devices that were previously not achievable in other single material platforms. Here, we exploit the electrical tunability of MATBG to engineer Josephson junctions and tunneling transistors all within one material, defined solely by electrostatic gates. Our multi-gated device geometry offers complete control over the Josephson junction, with the ability to independently tune the weak link, barriers, and tunneling electrodes.  
We show that these purely 2D MATBG Josephson junctions exhibit nonlocal electrodynamics in a magnetic field, in agreement with the Pearl theory for ultrathin superconductors \cite{pearl_current_1964}.
Utilizing the intrinsic bandgaps of MATBG, we also demonstrate 
monolithic edge tunneling spectroscopy within the same MATBG devices and measure the energy spectrum of MATBG in the superconducting phase.
Furthermore,  by inducing a double barrier geometry, the devices can be operated as a single-electron transistor, exhibiting Coulomb blockade. These MATBG tunneling devices, with versatile functionality encompassed within a single material, may find applications in graphene-based tunable superconducting qubits, on-chip superconducting circuits, and electromagnetic sensing in next-generation quantum nanoelectronics.
}

Tunneling devices are ubiquitous in modern electronics, with applications ranging from tunneling diodes to magnetic tunnel junctions and superconducting Josephson Junctions (JJ).
These devices typically involve heterojunctions of different materials to achieve conducting electrodes in series with a weak link or insulating barrier \cite{tinkham__michael_introduction_2004, Likharev_1979}.
In superconducting circuits, state-of-the-art JJs utilizing oxide tunnel barriers often suffer from disorder and localized states in the noncrystalline barriers \cite{oliver_materials_2013}.
Semiconductor-based JJs necessarily involve heterojunctions, and thus potentially non-ideal interfaces, but offer some advantages for integrated electronics, such as partial tunability of the semiconducting weak link \cite{larsen_semiconductor-nanowire-based_2015, wang_coherent_2019}.
While this offers a number of different operation regimes, the ability to independently tune the electrodes into different phases would enable an exponentially larger number of tunable configurations, qualitatively changing the nature of the device \textit{in situ}.
A superconducting junction made of a single clean material, which simultaneously offers a high degree of tunability both in the weak link and in the superconducting electrodes themselves, is therefore highly desirable, but has not been realized to date.

The recent discovery of correlated insulators and superconductivity in MATBG accessible via electrostatic doping \cite{cao_correlated_2018, cao_unconventional_2018, yankowitz_tuning_2019, lu_superconductors_2019} makes this possible, introducing MATBG as an unexplored material platform for superconducting electronics. In twisted bilayer graphene, a moiré pattern emerges from the coupling between two vertically-stacked graphene lattices with a relative twist angle \cite{li_observation_2010}. Near the first ‘magic-angle’, a nearly-flat electronic structure \cite{suarez_morell_flat_2010, bistritzer_moire_2011} leads to a large density of states and electron localization in real space around the AA-stacked regions \cite{lopes_dos_santos_continuum_2012, cao_correlated_2018, cao_unconventional_2018}, resulting in strong electronic interactions and emergent many-body correlated states. Using electrostatic gating, a plethora of possible regimes, including p-n junctions, superconducting regions, metallic leads, and insulating islands, among others, become accessible in a single device, making this system attractive both for scientific and technological applications. We exploit this unprecedented tunability to create an all-in-one device that can be used both for superconducting electronics and normal-state operations, bridging the fields of tunable van der Waals materials and superconducting circuits.
This could open the door towards fully integrated superconducting electronics, where entire circuits are made out of a single material with local gates and customizable coupling within and between each of the electronic components.

In this Letter, we demonstrate the versatility of multiply-gated MATBG devices.
We report on fully tunable lateral JJs, where both the superconducting electrodes and the weak link can be locally controlled. Such JJs additionally provide definitive evidence of superconducting phase coherence in MATBG. 
Independent control of the weak link is performed via applying a top gate voltage, achieving a junction that can be continuously switched from insulating, metallic, to superconducting regimes, generating a tunable critical supercurrent.
In the same multi-gated devices we obtain spectroscopic evidence of the superconducting gap in MATBG by utilizing its intrinsic bandgaps to create lateral tunneling barriers.
Finally, inducing barriers on either side of a narrow MATBG strip allows us to transform the device into a single-electron transistor displaying periodic Coulomb diamonds.

\begin{figure}
\includegraphics[width=\textwidth]{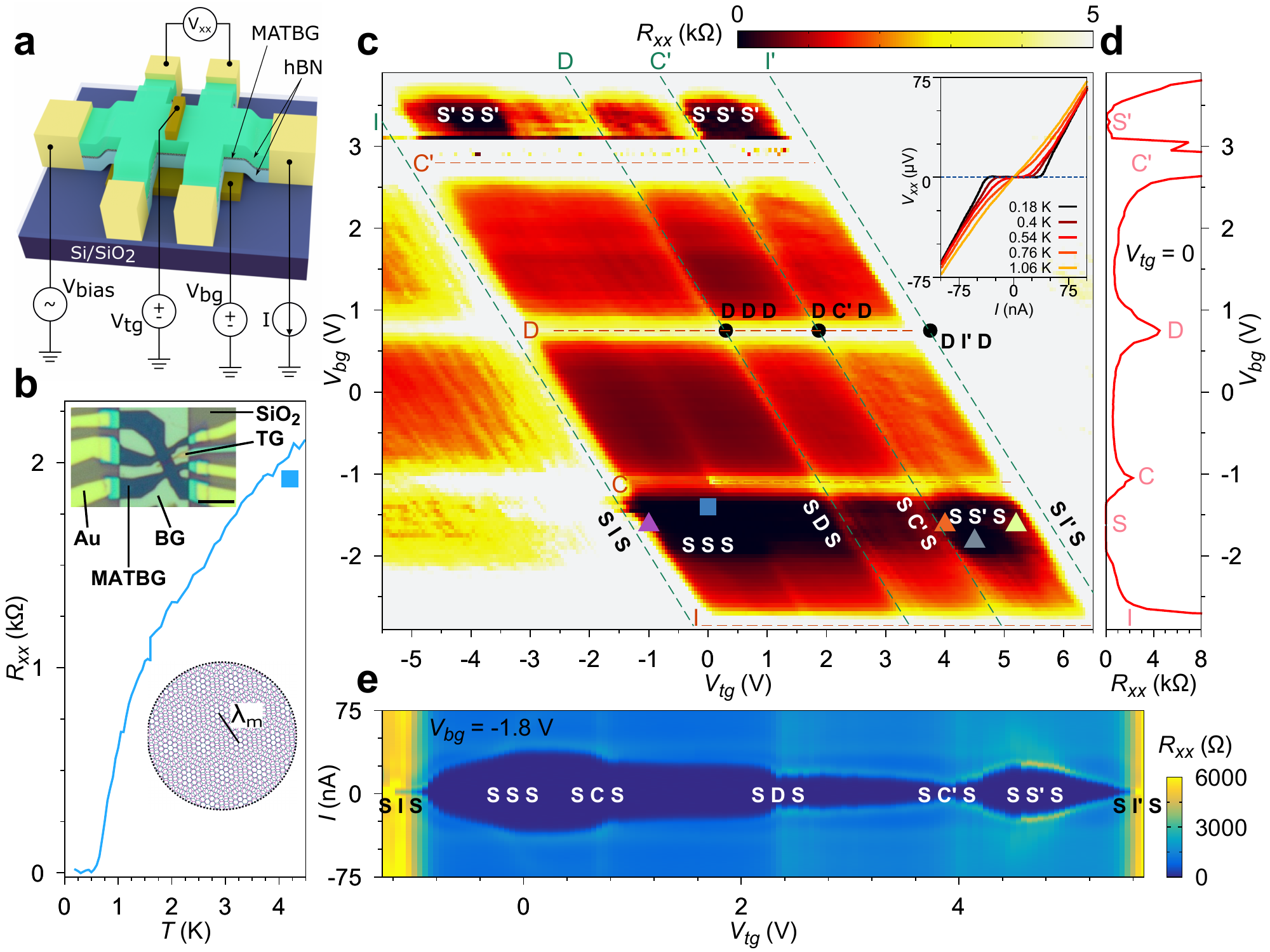}
\caption{\label{fig:fig1} Device 'A' structure and transport characterization . (a) Schematic illustration: A narrow top gate ($\sim \SI{160}{nm}$ wide) controls the electronic state of the region underneath. (b) Resistance vs. temperature curve measured at the blue square in panel (c)). Upper left inset: optical image of the final device. A back gate (BG) tunes the electron density in the overlapping region of the MATBG. The narrow top gate (TG) controlling the electronic state of the weak link can be seen at the center of the device. A bias voltage $V_{bias}$ is applied between the drain and source electrodes, and the 4-probe resistance $R_{xx} = V_{xx} /I$ is measured. The scale bar corresponds to $\SI{4}{\micro m}$.
Lower right inset: moiré pattern in twisted bilayer graphene. The displayed twist angle is enlarged for clarity with respect to the first magic angle $\theta \sim 1.1 ^\circ$. The moiré wavelength is given by $\lambda_m = a / [2 \sin(\theta/2)]$, where $a =  \SI{0.246}{nm}$  is the lattice constant of monolayer graphene, and $\theta$ is the twist angle. (c) Resistance as a function of the back gate and the top gate. The dark regions correspond to the superconducting states. Horizontal dashed lines and labels in red denote features induced by the back gate (see main text for explanation), and diagonal lines and labels in green denote features under the influence of both the top and the back gates. White labels of the form \textit{SXS} or \textit{S'XS'} indicate that at these points the source and drain are in S or S' state while the top-gated regions is in the X=I', C', D, S, S', C, or I state. Color-coded triangles indicate the points at which Fraunhofer patterns are taken in Fig. 2. Inset: Current-Voltage curves measured at $V_{bg}= \SI{-1.6}{V}$, $V_{tg}= \SI{0}{V}$ at different temperatures. (d) Line-cut in panel (c) along $V_{tg}=0$. I = Full filling band insulating state, S = Superconducting, C = Correlated Insulator at Half Filling, D = Dirac (charge neutrality) point. Primes denote positive fillings (i.e., electron doping). (e) Resistance as a function of current bias and top gate voltage, for $V_{bg} = \SI{-1.8}{V}$ in the superconducting state.
}
\end{figure}

To demonstrate these effects, we have measured three superconducting devices labeled A, B, and C. All devices were fabricated using the tear-and-stack dry-transfer technique \cite{kim_van_2016, cao_superlattice-induced_2016, cao_correlated_2018}, which enables us to achieve high-quality devices with clean interfaces and twist angles close to the first magic angle, $\theta \sim 1.1 ^\circ$ (see Supplementary Information for the details of the fabrication process). Here, we focus on device A with a twist angle 0.95$^\circ \pm$0.02$^\circ$ (see Supplementary Information for devices B and C). Fig. \ref{fig:fig1}a shows the dual-gated device structure. An optical micrograph of the device is shown in the inset of Fig. \ref{fig:fig1}b. The entire device is gated by the back gate, while the top gate is patterned into a narrow strip ($\sim \SI{160}{nm}$) at the center of the device to enable independent control of the region underneath it.
Fig. 1b shows the temperature dependence of the resistance of device A at the optimal doping with no top gate voltage applied, $V_{tg}= \SI{0}{V}$, and back gate $V_{bg}=\SI{-1.4}{V}$ (corresponding to the blue square in Fig. \ref{fig:fig1}c), displaying a superconducting transition at $T_c \sim  \SI{0.85}{K}$, as estimated from 50$\%$ of the normal state resistance (see Supplementary Information for devices B and C). The non-linear $I$-$V$ curves captured near optimal doping, shown in the inset of Fig. \ref{fig:fig1}c, display zero resistance up to a critical current $I_c\sim \SI{35}{nA}$.

To map out the complex electrostatic response of these devices, we now explore the complete dual-gate parameter space available, which exhibits a number of different insulating, metallic, and superconducting states.
Using the back gate and narrow top gate together, we can define three separate regions within the same device with independent phases in series.
In each region of the MATBG, when the four-fold valley and spin-degenerate bands are fully filled or fully depleted at densities $n=\pm n_s$, where $n_s=4/A$ and $A$ is the area of the moir\'e unit cell, the system behaves as a band insulator \cite{cao_superlattice-induced_2016, kim_tunable_2017, nam_lattice_2017}. In the following, we denote the insulator at $-n_s$ as I, and the insulator at $n_s$ as I'. Correlated insulator states are observed at $n = \pm n_s/2$, and we denote them as C ($-n_s/2$) and C' ($n_s/2$), respectively. Similarly, S and S' denote the superconducting states near $\mp n_s/2$, respectively. Let us also denote D as the charge neutrality (Dirac) point, and n (n’) the normal metallic states at fillings $n<-n_s$ ($n>n_s$), when the higher energy dispersive bands become populated by holes (electrons).
Metallic states are also observed throughout the flat bands, away from charge neutrality and the correlated fillings, denoted as N.
Fig.~\ref{fig:fig1}d shows a vertical line cut of the resistance map at $V_{tg} = \SI{0}{V}$. From top to bottom, the peaks correspond to the states C', D, C, and I, respectively, while the dips indicate the superconducting states S' and S. 
The transition between the different series combinations of the central and outside regions are readily seen from the horizontal and diagonal features of the $V_{tg}$-$V_{bg}$ resistance map shown in Fig. \ref{fig:fig1}c.
We interpret the diagonal features (dependent on both $V_{tg}$ and $V_{bg}$) as stemming from the dual-gated region beneath the top gate, and the horizontal features (independent of $V_{tg}$) as coming from the regions outside the top-gated area.
The intersection between a few horizontal (red) and diagonal (green) lines are labeled with black circles in Fig. \ref{fig:fig1}c. For example, DDD denotes the coincidence of the Dirac points in all three regions, whereas DC'D occurs when the central region enters the C' correlated insulator state and the outside regions are at charge neutrality, and similarly for other intersections of the dashed lines.
More interesting device behavior is obtained by doping away from the horizontal lines in the dual-gate map.
For instance, supercurrent through a variable Josephson junction is observed across the device if we use $V_{bg}$ to globally tune the MATBG into S and use $V_{tg}$ to form a weak link in the central region with another state such as I, I', or C' (as indicated by the diagonal labels in Fig.~\ref{fig:fig1}c).
Figure \ref{fig:fig1}e illustrates the wide region of supercurrent observed (dark blue), whereas the ability to continuously vary the barrier strength with $V_{tg}$ is indicated by the evolution of the critical current (where the differential resistance becomes finite).
We can turn off the supercurrent completely by gating the central region deep into the insulating state (beyond full filling).
In this regime the superconducting coherence across the junction is lost, and a dissipative junction is obtained.

\begin{figure}
\includegraphics[width=\textwidth]{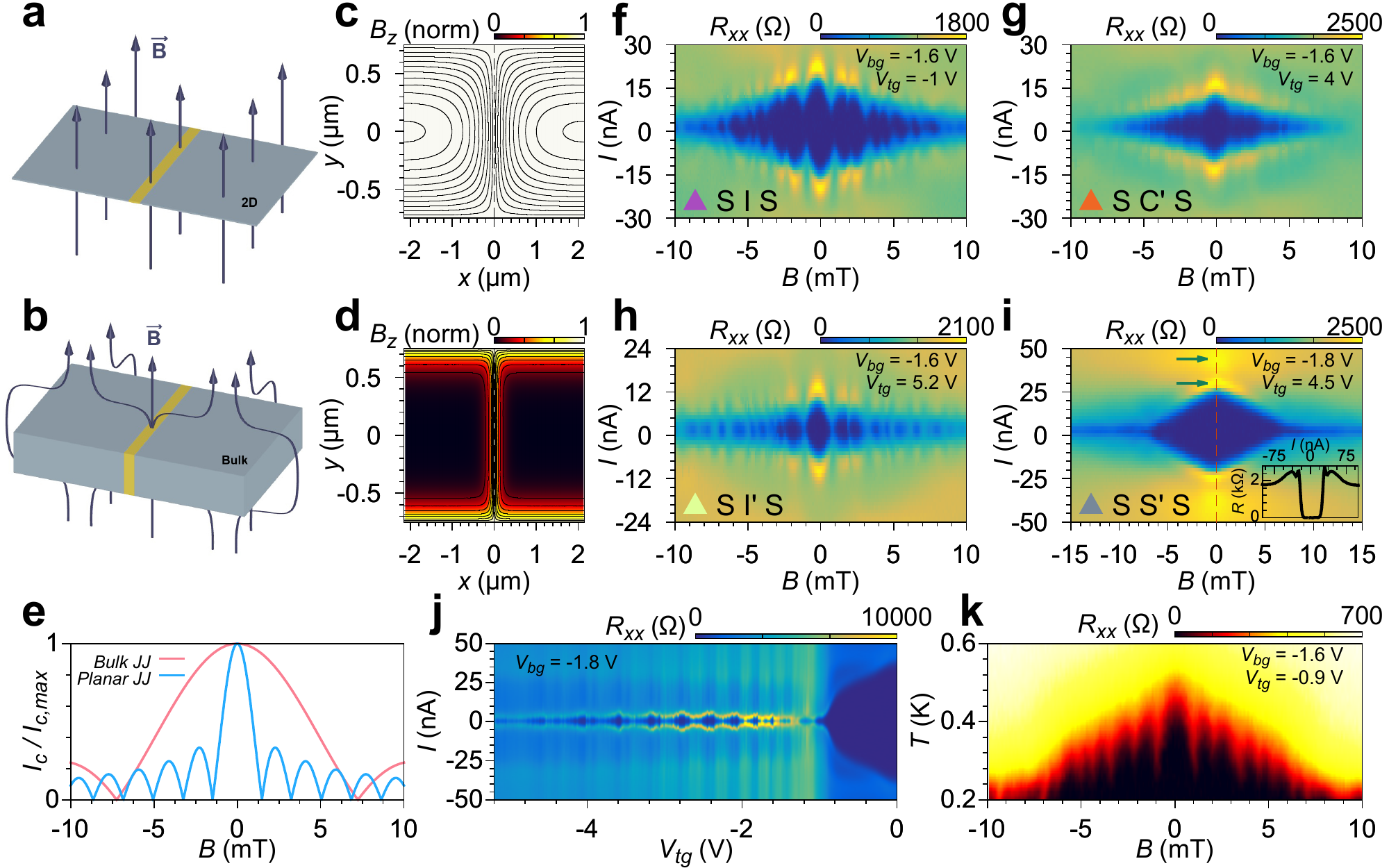}
\caption{\label{fig:fig2} Nonlocality and tunability of MATBG JJs. (a-b) Schematic representation of (a) a planar 2D Josephson junction and (b) a bulk Josephson junction in an external magnetic field. (c-d) Simulated distribution of the normalized magnetic field $B_{z}$ in a Josephson junction located at $x=0$ in the case of (c) planar and (d) bulk superconductor. The streamlines denote the flow of screening currents in the superconductor. The bulk case assumes a penetration depth of $\lambda = \SI{100}{nm}$. (e) Calculated critical current (normalized) as a function of the magnetic field for bulk and planar Josephson junctions. 
(f-i) Measured Fraunhofer pattern in device A (f) close to SIS regime, (g) SC'S regime, (h) SI'S regime, and (i) SS'S regime. Color-coded triangles correspond to those in Fig. \ref{fig:fig1}c.
Inset in (i) shows a linecut across the red dashed line at $B=0$ T. The two pairs of peaks in the curve correspond to the critical currents of the S and S’ states, respectively (green arrows in main panel). (j) Fabry-Pérot-like oscillations in the critical current. (k) Resistance as a function of magnetic field and temperature of the junction close to the SIS regime.}
\end{figure}

Next we address the expected behavior for 2D JJs in the presence of magnetic flux.
2D superconductors screen external magnetic fields in a fundamentally different way from their bulk (3D) counterparts. In ultra-thin superconductors where the film thickness is less than the London penetration depth $\lambda$, the characteristic length that governs the spatial magnetic field distribution is given by the Pearl length \cite{pearl_current_1964} $\Lambda = 2\lambda^2 / d \gg \lambda$, where $d$ is the film thickness. In the case of MATBG, the thickness is less than $\SI{1}{nm}$ and the Pearl length can reach macroscopic dimensions, exceeding the dimensions of the device itself. Under such conditions, the screening currents cannot effectively expel the external magnetic field (illustrated in Fig. \ref{fig:fig2}a), in striking contrast with bulk samples (Fig. \ref{fig:fig2}b). The origin of this effect can be understood by recognizing that the self-field of the screening current in a thin-film superconductor scales as $\frac{w}{\Lambda}B$, where $w$ is the lateral dimension of the sample and $B$ is the external magnetic field \cite{moshe_edge-type_2008}. When $w\ll \Lambda$, the self-field is therefore negligible compared to the external field, allowing finite field penetration.

The distribution of Josephson current and magnetic flux in a JJ is altered in the 2D limit as well. In a bulk JJ, the magnetic field only penetrates a distance $\sim\lambda$ into the superconductor and is therefore mostly confined within the junction barrier. The phase difference across the junction is a simple function that only depends on $\lambda$ and $B$. On the other hand, for edge-type Josephson junctions in ultra-thin films, the magnetic field distribution is not confined to the tunneling barrier.
An additional distinction between bulk and edge-type 2D JJs is that the Josephson electrodynamics are nonlocal in the 2D case, that is, the magnetic flux in the junction results from a non-negligible superconducting phase gradient in both 2D superconducting regions\cite{ivanchenko_nonlocal_1990, boris_evidence_2013, tinkham__michael_introduction_2004, jr_nonlocal_2009}.

To further understand the physical implications of these differences, we simulate the magnetic field and the screening currents in a bulk JJ and a 2D JJ with similar dimensions as device A, placed in an external magnetic field $\textbf{B} = B_0 \hat{z} $\cite{clem_josephson_2010, kogan_josephson_2001, moshe_edge-type_2008} (see Supplementary Information for details). 
While in bulk superconductors, the magnetic field decays exponentially at the edges within a distance $\lambda$ (Fig.~\ref{fig:fig2}d), in the 2D case, it is essentially unaltered across the entire sample area (Fig.~\ref{fig:fig2}c). From the magnetic field distribution, we can obtain the distribution of the phase difference across the junction which gives rise to the Fraunhofer interference pattern (see Supplementary Information). Our calculated field dependence of the maximum supercurrent in the 2D JJ, in agreement with previous analytical and numerical predictions \cite{clem_josephson_2010, kogan_josephson_2001, moshe_edge-type_2008, rosenthal_flux_1991}, differs noticeably from the typical Fraunhofer pattern of bulk junctions in two ways (Fig. \ref{fig:fig2}e).
First, the high-field periodicity $\sim 1.8 \Phi_0/w^2$ depends solely on the geometry of the sample and is usually much smaller than the bulk periodicity $\Phi_0 / w(a+2 \lambda)$, where $a$ is the length of the weak link itself. Second, unlike the bulk case, the zeros of $I_{c}(B)$ for edge-type thin-film junctions are not equidistant at low-fields.

We now present measurements of Josephson behavior in a perpendicular magnetic field to test these predictions experimentally. Based on the analytical expression $1.8 \Phi_0/w^2$, we expect an interference period $\sim \SI{1.7}{mT}$ in device A ($w \sim \SI{1.5}{\micro m}$). We first gate the device into the SIS regime, pushing the middle region as far as possible into the insulating region while maintaining superconducting coherence across the junction (Fig. \ref{fig:fig2}f).
We observe oscillations in the critical current with a periodicity of $\sim \SIrange{1}{1.5}{mT}$ for the finest oscillations.
If one only considers the physical length and width of the junction, an approximation of the bulk formula $\Phi_0/wl$ using $l\approx a$ gives a periodicity of  \SI{8.5}{mT}, significantly larger than the measured oscillation period. However, our measured periodicity of $\sim \SIrange{1}{1.5}{mT}$ is clearly in agreement with the simulations shown in Fig.~\ref{fig:fig2}e and consistent with the Pearl regime governing the ultrathin superconducting electrodes. This anomalous periodicity is further corroborated by two additional devices (B and C), shown in the Supplementary Information. Critical current oscillations with similar periodicity are also observed close to the SC'S (Fig. \ref{fig:fig2}g) and SI'S (Fig. \ref{fig:fig2}h) configurations. Although there are slight deviations from ideal Fraunhofer behavior and differences between each pattern, these details may be attributed to inhomogeneities and asymmetries across each junction. Overall, SIS-type junction behavior is clearly achievable in this geometry using different insulating phases. Alternatively, if we bring the weak link into the SS’S regime, the oscillations in the critical current disappear (Fig. \ref{fig:fig2}i). Notably, in this arrangement, there is a complex spatial variation of the carrier density across the junction, which continuously traverses the phase diagram from the S state to the S' state and back to the S state. Despite this spatial density profile, the junction simply behaves like 3 superconductors in series, where the global critical current is determined by the smaller critical current of the S' region. The lack of oscillations and the presence of two clear critical current peaks in the $dV_{xx}/dI$ at zero field (green arrows in Fig.~\ref{fig:fig2}i) supports this conclusion. 
In addition to the anomalous periodicity of the measured Fraunhofer patterns, another piece of evidence for the nonlocal Josephson effect in our samples \cite{boris_evidence_2013} is the resistance as a function of the magnetic field and temperature of the junction close to the SIS regime (Fig.~\ref{fig:fig2}k). The oscillation period does not change as the temperature approaches $T_c$. This is in contrast to the expected behavior in the local regime where, since $\lambda(T)$ diverges as $T\rightarrow T_c$, one expects the oscillation period, which is inversely proportional to $\lambda$, to be progressively suppressed with temperature \cite{nagata_oscillations_1982}.

In the SnS regime, we find oscillatory behavior of the critical current with respect to $V_{tg}$, as shown in Fig. \ref{fig:fig2}j. We attribute this to a Fabry-Pérot-like resonance from the interfaces between the dual-gated region and the singly gated regions, which occurs in high-quality devices close to the ballistic transport regime \cite{calado_ballistic_2015, ben_shalom_quantum_2016}. In a JJ, $I_c$ and the normal state resistance $R_N$ typically scale inversely with each other, a behavior known as the Ambegaokar-Baratoff relationship \cite{tinkham__michael_introduction_2004}. When the Fabry-Pérot resonance of the electron wave becomes prominent, $R_N$ is periodically modulated by $\sqrt{n+n_s}$, and so is the critical current in the opposite way. We observe the resonance only in the SnS regime, likely due to the low effective mass and high mobility in the dispersive band at $-|n|< -|n_s|$.

\begin{figure}
\includegraphics[width=\textwidth]{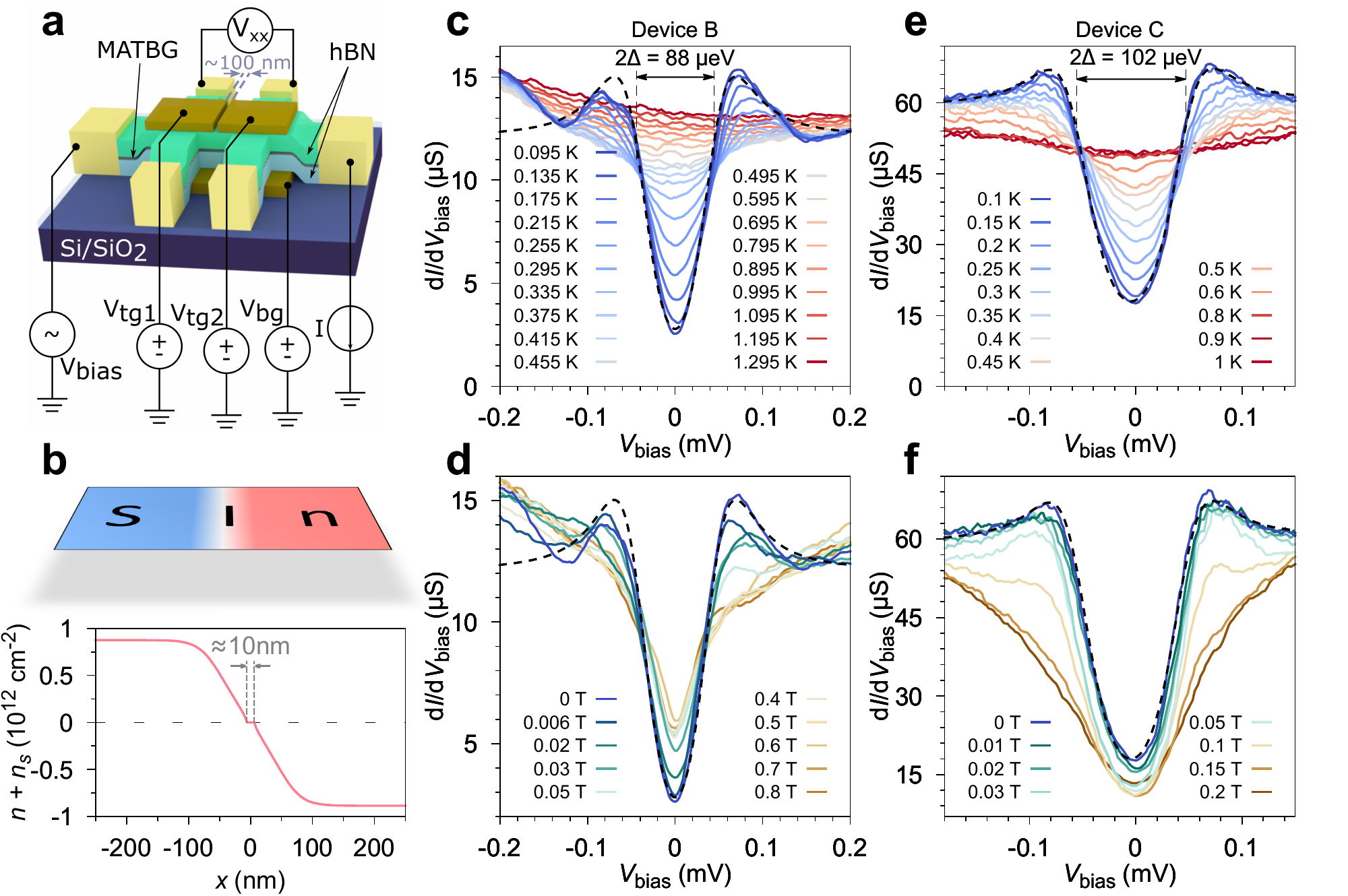}
\caption{\label{fig:fig3} 
Edge tunneling spectroscopy of the superconducting gap in MATBG. (a) Structure of devices B and C with two top gates separated by a narrow gap. (b) Numerical calculated charge carrier density distribution in the p-n junction regime performed for device B (device C analogous). The left half is brought into the superconducting state S at $n \sim - 0.6 |n_s|$, while the right half is brought into a normal metallic regime with density $n \sim - 1.4 |n_s|$ (all densities are for hole doping and thus are all negative). In the central region of the device, the density passes through the band insulator I ($n = -n_s$), thus creating a tunneling barrier. (c) Raw tunneling spectra as a function of temperature, from $\SI{0.095}{K}$ to $\SI{1.295}{K}$  for device B. The black dashed line is a theoretical fit to the quasiparticle density of states (see Supplementary Information), yielding an extracted gap $\Delta_\text{fit} = \SI{44}{\micro eV}$ (with negligible broadening). (d) Magnetic field dependence of the edge tunneling spectra, from 0~T to 0.8~T for device B. (e) Similarly, tunneling spectra as a function of temperature and (f) perpendicular magnetic field for device C, with the left half of the device in the superconducting state S at $n \sim - 0.775 |n_s|$, and the right half at a density $n \sim - 1.2 |n_s|$. The dashed line is the theoretical fit giving an extracted corresponding gap $\Delta_\text{fit} = \SI{51}{\micro eV}$ (with Dynes broadening $\SI{15}{\micro eV}$).
}
\end{figure}

We now turn our attention to devices B and C with the structure shown in Fig.~\ref{fig:fig3}a. Instead of the narrow top gate in device A, here we pattern two isolated top gates separated by a narrow gap, allowing us to realize a p-n junction. In Fig. \ref{fig:fig3}b, we show the simulated charge carrier density distribution across a gate-defined p-n junction in a scenario similar to device B (see Supplementary Information for simulation details). The density evolves continuously and crosses the value $n = -n_s$ at a position between the left gated and right gated regions. Due to quantum capacitance effects in MATBG, a narrow region in the p-n junction is kept inside the bandgap at $-n_s$ and acts as a tunneling barrier. If we put one side of the junction in the S state, we then realize an nIS configuration, enabling edge tunneling spectroscopy into the S state.
Using this configuration, in Fig. \ref{fig:fig3}c-f we show tunneling spectra of MATBG in the superconducting regime.

The data show clear spectroscopic evidence of a superconducting gap, including well-defined coherence peaks and a minimum at zero bias.
To obtain a quantitative measure of the gap that incorporates the entire spectral lineshape, we fit the data to a model for the quasiparticle density of states.
Choosing the simplest such model, that of a conventional isotropic s-wave order parameter \cite{tinkham__michael_introduction_2004}, we incorporate the effects of thermal and lifetime (Dynes) broadening, and extract a gap of $\Delta_\text{fit} = \SI{44}{\micro eV}$ at $T =  \SI{95}{mK}$ for device B, and $\Delta_\text{fit} = \SI{51}{\micro eV}$ at $T = \SI{100}{mK}$ for device C (see Supplementary Information for fitting details). The tunneling conductance minimum and coherence peaks are well captured by this fit, including the absence of a hard gap due to thermal broadening at the lowest experimental temperature. Taking the $\Delta_\text{fit} = \SI{51}{\micro eV}$ for device C as a lower bound for the superconducting gap at zero temperature ($\Delta_\text{fit} \lesssim \Delta_0$), we can estimate an associated transition temperature from the BCS approximation $T_c \gtrsim \Delta_\text{fit} / (1.764 k_B) \approx \SI{340}{mK}$. This value is reasonable considering the transition temperature extracted from 50$\%$ of the normal-state resistance at a nearby doping value, $\sim \SI{400}{mK}$ (see Supplementary Information). However, we emphasize that such a fitting procedure cannot distinguish the symmetry of the superconducting order parameter in our data, as there is significant spectral broadening due to temperature, disorder, and the lateral junction geometry. In fact, we have found equally good quality fits using other non s-wave order parameters. Direct measurements of the pairing symmetry in MATBG remain a fundamental yet unresolved question in the field, and these device structures may be adapted to shed light on this topic.

As the temperature is increased above $\sim \SI{300}{mK}$ (for both devices B and C), the coherence peaks are significantly broadened due to thermal excitations.
As the temperature is further increased, the dip at $V_\mathrm{bias}=0$ is suppressed and eventually disappears, indicating that the system is no longer superconducting. 
Similarly, by applying a perpendicular magnetic field at base temperature, the coherence peaks are also suppressed at $B \sim  \SI{50}{mT} $ for device B ($B \sim  \SI{100}{mT}$ for device C), comparable to the upper critical field observed in transport in magic-angle devices with similar $T_c$ \cite{cao_unconventional_2018}.
The closing of the gap and suppression of the coherence peaks with temperature and magnetic field further support a superconducting origin for the observed gap.
Notably, a portion of the tunneling minimum at zero bias persists to much larger magnetic fields. 
However, a similar tunneling minimum is observed above the critical field for a wide range of densities outside of the superconducting dome, without associated coherence peaks at zero field (see Supplementary Information), and thus arises from a distinct mechanism. 
Such a suppression in the tunneling spectra may be related to the Efros-Shklovskii-type Coulomb gap that arises at the Fermi level due to localization in disordered semiconducting thin films \cite{efros_coulomb_1975, lee_coulomb_1999}, or alternatively, it may result from electronic interactions during the tunneling process \cite{b.l.altshuler_zero_1993, gershenzon_tunnel_1986, kotelnikov_zero-bias_2013}. Although we consistently observe a suppression in the tunneling conductance at zero bias for all three measured devices (see Supplementary Information), further detailed studies are required to determine the precise origin of this spectral feature.

\begin{figure}
\includegraphics[width=\textwidth]{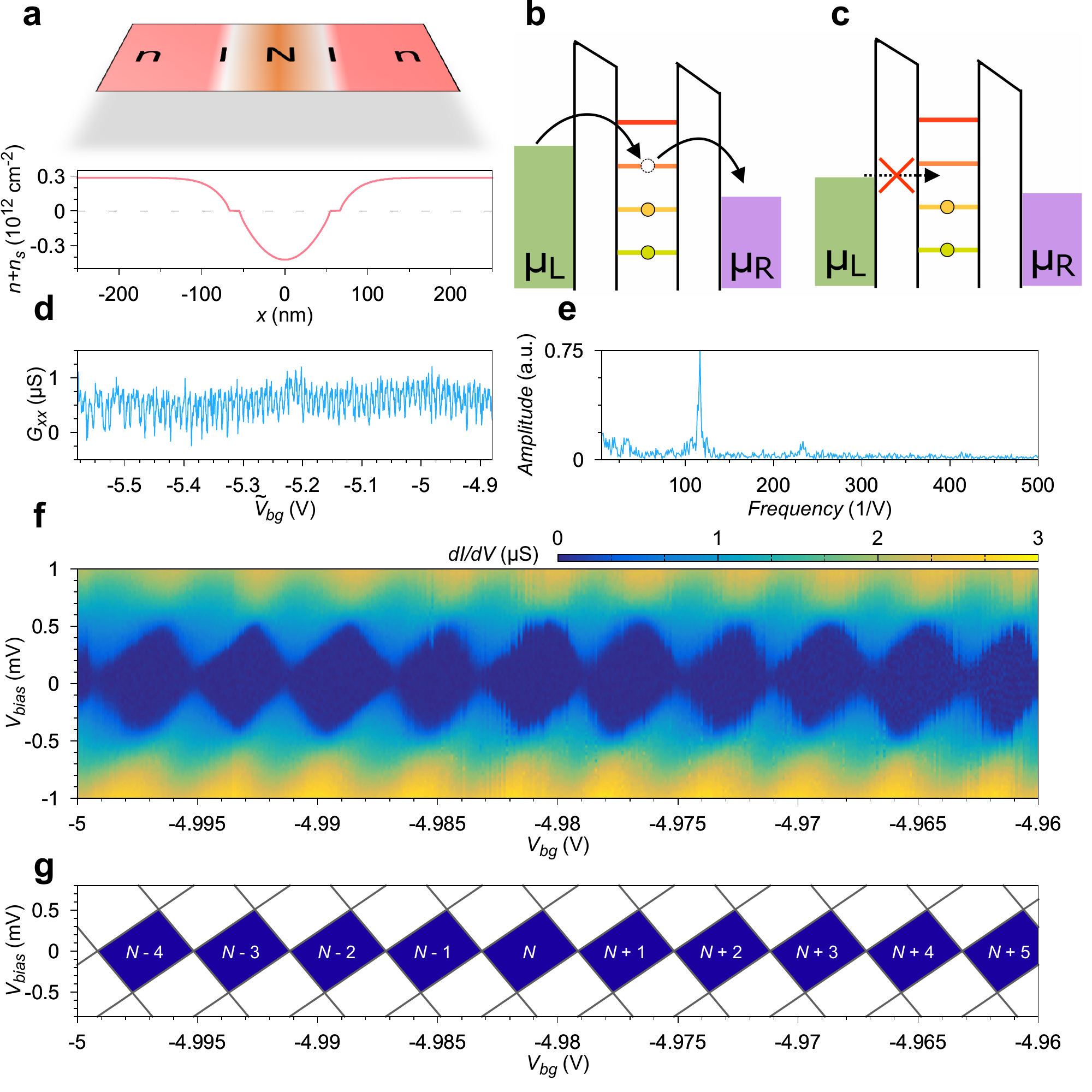}
\caption{\label{fig:fig4} 
\emph{In situ} single-electron transistor and Coulomb blockade in MATBG. (a) Gating scheme of the device and simulation of the charge density across the sample. Two effective insulating barriers emerge, and the central island is equivalent to a single-electron transistor. (b-c) Space-energy diagram of a single-electron transistor in the (b) non-blocked and (c) blocked regimes. $\mu_L$ and $\mu_R$ are the chemical potentials of the source and the drain, respectively. (d) Conductance versus the back gate voltage $V_{bg}$, while the two top gates keep the densities on the source and drain at $n \sim -1.1 n_s$. (e) Fourier transform of the two-probe tunneling current, showing a single peak at $\SI{116}{V^{-1}}$. (f) Differential conductance as a function of back gate voltage and source-drain bias voltage. Pronounced Coulomb diamonds, corresponding to the absence of tunneling current, are observed. In this scan, only the back gate is swept while both top gates are fixed. (g) Schematic of the Coulomb diamonds (see main text). $N$ denotes the number of electrons in the central island. The mismatch of the periodicities in panels (d-e) and panels (f-g) is attributed to cross-coupling of the top gates and is discussed in the Supplementary Information (we have added a tilde in the x-axis label of panel (d) to avoid possible confusion). All the data in this figure correspond to device B (see Supplementary Information for additional data).
}
\end{figure}

Further exploiting the flexibility of the split-gate geometry, we can create a single-electron transistor (SET) within the same multipurpose devices. In a SET, electrons are spatially confined in a central region by tunneling barriers that weakly connect it to the drain and source electrodes. To achieve this, we tune the left and right top gates to bring the two sides of the device into metallic states with densities $n < -n_s$. The central narrow region, being singly gated, is brought into the density range $-n_s < n < -n_s/2$. With similar arguments as those mentioned above, two insulating plateaus with $n=-n_s$ form around the central region of the sample, resulting in an isolated island in the middle. Fig. \ref{fig:fig4}a illustrates such an nINIn configuration, as seen from above, and the calculated charge carrier density distribution. At low temperatures ($k_BT \ll e^2/C$, where $C$ is the total capacitance of the island) and with large tunneling resistance ($R_\mathrm{tunnel} \gg h/e^2$), electron tunneling is allowed only if there are available discrete energy levels between the Fermi energies of the source and the drain (Fig. \ref{fig:fig4}b), whereas the Coulomb blockade effect prohibits tunneling otherwise (Fig. \ref{fig:fig4}c) \cite{tinkham__michael_introduction_2004, ihn_graphene_2010}.
 
In Fig.~\ref{fig:fig4}d, we measure the tunneling conductance of the gate-defined single-electron transistor in device B as the back gate voltage is varied, keeping the source and drain densities fixed. The signal displays fine, reproducible oscillations as a function of the back gate voltage.
A Fourier transform of the measured tunneling current reveals a single periodicity, as shown in Fig.~\ref{fig:fig4}e. Fig.~\ref{fig:fig4}f shows the differential conductance in a narrower range of back gate voltages, as a function of the source-drain bias voltage. We observe well-developed Coulomb diamonds with zero conductance in the blockaded regime. 
These observations are in agreement with a single-electron transistor with capacitances $C_g = \SI{40}{aF}$, $C_1 \approx C_2 = \SI{110}{aF}$, $C_{\Sigma} = \SI{310}{aF}$ (see Supplementary Information for definitions), as shown in Fig. \ref{fig:fig4}g for a charged island with an integer number of electrons on the island, labeled by $N$. 
Thus we find that the band insulator in MATBG provides a suitable barrier for SET physics in graphene with appropriate local gating, adding to the broad tunability of these MATBG devices. 

The unprecedented tunability of MATBG together with local electrostatic gating in this work enables complete control of the weak link and junction electrodes, independently.
With this versatile platform, we demonstrate multiple Josephson junctions with differing barrier strength and character, edge-tunneling spectroscopy of the superconducting state, and robust SET physics in a double-barrier configuration.
While multiple devices are presented here, critically, all three aforementioned experiments are achievable in a single device geometry.
Gate-defined tunnel junctions present a significant advance toward probing the superconducting order parameter in MATBG, and will inspire further advances for exploring physics within the expanding class of moir{\'e} systems.
Furthermore, these multipurpose devices establish a clear path toward gate-defined circuits with MATBG in future 2D integrated electronics, with potential applications in low-temperature circuits, quantum computing, and electromagnetic sensing.

\bibliographystyle{Nature}
\bibliography{references} 

\end{document}